\documentclass[lettersize,journal]{IEEEtran}
\usepackage{hyperref}
\usepackage{color,soul}
\usepackage{graphicx}
\usepackage[font=scriptsize]{caption}
\usepackage{subcaption}
\usepackage{booktabs}
\usepackage{siunitx}
\usepackage{enumitem}
\usepackage{amsmath}
\usepackage{amssymb}
\usepackage{multirow}
\usepackage{xurl}
\usepackage{float}
\usepackage[british]{babel}
\usepackage{csquotes}
\usepackage[mathlines]{lineno}
\usepackage{xcolor,soul}
\sethlcolor{yellow}
\usepackage{algorithm}
\usepackage{algpseudocode}
\usepackage{orcidlink}

\makeatletter
\renewcommand{\fnum@figure}{Fig. \arabic{figure}. }
\renewcommand{\fnum@table}{TABLE \roman{table}}
\makeatother
\begin{document}
\title{Attitude and Heading Estimation in Symmetrical Inertial Arrays}
\author{Yaakov~Libero \orcidlink{0000-0001-5213-9985}, and Itzik Klein \orcidlink{0000-0001-7846-0654}%
\thanks{Yaakov Libero and Itzik Klein are with the Hatter Department of Marine Technologies, Charney School of Marine Sciences, University of Haifa, Haifa, Israel.}
}
%
\maketitle
\begin{abstract}
Attitude and heading reference systems (AHRS) play a central role in autonomous navigation systems on land, air and maritime platforms. AHRS utilize inertial sensor measurements to estimate platform orientation. In recent years, there has been increasing interest in multiple inertial measurement units (MIMU) arrays to improve navigation accuracy and robustness. A particularly challenging MIMU implementation is the gyro-free (GF) configuration, in which angular velocity is derived solely from accelerometer measurements. While the GF configurations have multiple benefits, including outlier detection and in angular acceleration measurements, their main drawbacks are inherent instability and an increased divergence rate. To address these shortcomings, we introduce a novel symmetrical MIMU formulation, in which the IMUs are arranged in symmetric diagonal pairs to decouple linear and rotational acceleration components. To this end, we derive the theoretical foundations for the symmetrical MIMU formulation of the GF equations, develop a nonlinear least squares estimation process, and integrate statistical hypothesis testing into an AHRS error-state extended Kalman filter. We validate our approach using real-world datasets containing 85 minutes of navigation data recorded on both airborne and land platforms. Our results demonstrated a 30\% average reduction in attitude estimation errors, rotation detection accuracy exceeding 95\% improvement, and significantly improved stability compared to a standard GF implementation. These results enable reliable GF navigation in applications where gyroscopes are unavailable, unreliable, or energy-constrained. Common examples include miniature platforms, computational-constraint platforms, and long-endurance marine platforms.
\end{abstract}
\IEEEpeerreviewmaketitle
\section{Introduction}
Attitude and heading reference systems (AHRS) are used in a wide range of applications, including autonomous vehicles \cite{sabet2017experimental}, robotics \cite{wang2014simple}, avionics \cite{sipos2009flight}, and marine navigation \cite{kim2011study}. AHRS systems provide orientation in environments relaying solely on internal sensors \cite{vertzberger2022adaptive, guerrero2011design, farrell2008aided}. This makes them a central component in real-world applications where external updates may be unreliable or unavailable \cite{kwon2020performance}.

In recent years, the increased availability of low-cost inertial measurement units (IMUs) has enabled the development of multiple-IMU (MIMU) arrays \cite{patel2022multi}. MIMU improve navigation performance and accuracy by reducing measurement noise through averaging processes, enabling outlier detection algorithms, and enhancing system robustness through IMU-redundancy \cite{guerrier2009improving}. However, effectively integrating MIMU presents multiple challenges, including handling sensor biases, increased implementation difficulties, predominantly in the design of fusion algorithms \cite{rasoulzadeh2016implementation}, and system stability \cite{bancroft2011data}.

MIMU array filtering is divided into two common approaches: virtualization and federalization \cite{bancroft2010multiple}. Virtual IMU (VIMU) \cite{huang2022optimal,patel2021sensor,fu2023adaptive} approaches handle multiple IMUs by fusing them into a single virtual IMU. While these approaches provide a net improvement in accuracy in the short term, they face challenges with sensor biases mitigation, which is essential when working with low-cost inertial sensors \cite{park2008error}. By contrast, Federated filters \cite{yang2021multilayer,bancroft2009multiple,mounier2025multi} work by implementing an individual filter, typically an extended Kalman filter, for each IMU and an averaging process, typically a least squares process, to average the different results into a unified solution. While they allow for a dynamic number of IMU and handle each IMU bias individually, they usually lead to subpar error reduction compared to VIMU approaches \cite{libero2024augmented}. 

A variation of the MIMU setup is the gyro-free (GF) configuration, in which angular velocity is derived from accelerometer measurements alone \cite{zhou2021gyro}. The GF configurations offer multiple advantages compared to traditional 3-axis accelerometer/gyro configurations. Those include reduced costs \cite{schopp2010design}, the ability to measure angular acceleration \cite{padgaonkar1975measurement} and additional benefits inherent to multi-sensor configurations, as seen in other MIMU arrays. However, the GF configurations come with some inherent limitations. They suffer from increased error growth rate and stability issues due to the double integration and autocorrelation included in their mechanization process \cite{nusbaum2017control}. GF can be used for various navigation applications including AHRS, yet the literature on GF-AHRS implementation is sparse. Work such as \cite{hanson2005using} includes the building blocks for an AHRS implementation but maintains a direct mechanization approach instead of a filter-based one. The reason the traditional Kalman filter-based AHRS implementation has proven challenging to implement in GF AHRS is that the Kalman filter assumption of independent process and measurement noise covariance \cite{bar2001estimation} does not hold as both depend on inertial sensors.

In this paper, we introduce a novel approach to GF AHRS by deriving a theoretical framework based on the properties of symmetric MIMU arrays. We also incorporate statistical hypothesis testing into the nonlinear estimation framework.

The contributions of this paper are:

\begin{enumerate}
    \item The development of a novel array configuration – the symmetrical MIMU (SMIMU) array, in which IMUs are arranged in symmetric diagonal pairs that enable the separation of linear and rotational acceleration components. We distinguish between 2D and 3D SMIMU configurations and derive the theory for both cases.
    \item Derive a comprehensive theoretical framework for the symmetrical IMU derivation of the GF equations and implement a matching nonlinear least squares estimation process for the GF IMU states.
    \item Integrating statistical hypothesis testing within an AHRS error-state extended Kalman filter to detect and reject noise-based propagation and thereby enable a more accurate state estimation.
\end{enumerate}

To demonstrate our approach, it was evaluated using a dataset of 85 minutes of navigation data collected from both airborne and land platforms. We show that significant improvement was achieved compared to the standard GF implementation. This included an improvement of 30\% in attitude estimation error, improved robustness compared to GF implementations, and rotation detection with an improvement of 95\% in accuracy. These results enable reliable GF navigation in applications where gyroscopes are unavailable, unreliable, or energy-constrained. Common examples include miniature platforms, medical robotics, computational-constraint platforms, and long-endurance marine platforms (e.g. buoys).

The remainder of this paper is organized as follows: Section II presents the problem formulation, including GF IMU equations, kinematic equations, and the conventional error-state AHRS Kalman Filter. Section III gives our proposed approach, encompassing symmetrical MIMU derivation, non-linear least squares estimation, nonlinear filter, and hypothesis testing methodology. Section IV presents the experimental validation using real-world datasets, followed by conclusions and future research directions provided in Section V.

\section{Problem Formulation}
\subsection{Gyro-Free IMU}
Consider an inertial reference frame and a moving and rotating rigid body frame, scattered with $\textrm{N}$ IMUs on arbitrary points along the body $\mathbf{p}_i\,|\,i\in[0,\textrm{N}]$. In the general case, the accelerometer output $\mathbf{\tilde{f}}_{s_i}$ in its sensor frame $s_i$ can be formulated by:

\begin{equation} 
    \label{eq:f_s}
    \mathbf{\tilde{f}}_{s_i} = \mathbf{a}_{s_i}^i - \mathbf{g}_{s_i}
\end{equation}

 Where $\mathbf{a}_{s_i}^{i}$ is the acceleration of point $i$, located at $\mathbf{p}_i$, with respect to an inertial frame, and $\mathbf{g}_{s_i}$ is the gravitational acceleration, both in the sensor reference frame.

Since the body frame is rigid, meaning the body frame can move and rotate in inertial space but points in the body frame will remain stationary relative to each other, the rotation between the arbitrary sensor and body frames $\mathbf{R}_{s_i}^b$ is constant. We can therefore derive an arbitrary accelerometers output in the body frame to be:

\begin{equation}
    \label{eq:f_b}
    \mathbf{\hat{f}}_{b} = \mathbf{a}_{b}^i - \mathbf{g}_{b}
\end{equation}

 With $\mathbf{a}^i_b$ and $\mathbf{g}_b$ being the i'th sensor and gravitational acceleration in the body frame, respectively.

Since the body frame is a rotating frame of reference, meaning it is no longer an inertial frame of reference, we cannot use Newton's second law of motion directly. To handle this, we will start by defining $\boldsymbol{\rho}_i$ to be the distance between the i'th point's position $\mathbf{p}_i$ and the body's center of mass $\mathbf{p}_b$

\begin{equation}
    \boldsymbol{\rho}_i = \mathbf{p}_i - \mathbf{p}_b
\end{equation}

 Assuming the center of mass is the origin of the body reference frame, we will use $\boldsymbol{\rho}_i$ to break down $\mathbf{a}_b^i$ into classical (inertial) and rotating (non-inertial) accelerations, with respect to an inertial reference frame, using fictitious forces, as demonstrated in \cite{hanson2005optimal}, \cite{Gregory_2006}:

\begin{equation}
    \label{eq:a_b_full}
    \mathbf{a}_b^i = \mathbf{a}_b + 2\boldsymbol{\omega}_{ib}^b\times\dot{\boldsymbol{\rho}}_i + \boldsymbol{\omega}_{ib}^b\times(\boldsymbol{\omega}_{ib}^b\times\boldsymbol{\rho}_i) + \boldsymbol{\dot{\omega}}_{ib}^b\times\boldsymbol{\rho}_i +\boldsymbol{\ddot{\rho}}_i
\end{equation}

 Where $\mathbf{a}_b$ is the body's acceleration relative to the inertial frame, $\boldsymbol{\omega}_{ib}^b$ and $\boldsymbol{\dot{\omega}}_{ib}^b$ are the angular velocity and acceleration, and $\boldsymbol{\dot{\rho}}_i$, $\boldsymbol{\ddot{\rho}}_i$ are the point's velocity and acceleration with respect to the body. 

However, since we established that the body is rigid, we can show that $\boldsymbol{\dot{\rho}}_i$ and $\boldsymbol{\ddot{\rho}}_i$ are equal to zero, simplifying (\ref{eq:a_b_full}) to:

\begin{equation}
    \label{eq:a_b_simp}
    \mathbf{a}_b^i = \mathbf{a}_b + \boldsymbol{\omega}_{ib}^b\times(\boldsymbol{\omega}_{ib}^b\times\boldsymbol{\rho}_i) + \boldsymbol{\dot{\omega}}_{ib}^b\times\boldsymbol{\rho}_i
\end{equation}

 Substituting (\ref{eq:a_b_simp}) into (\ref{eq:f_b}) results in:

\begin{equation}
    \label{eq:gf}
    \mathbf{\hat{f}}_{b} = \mathbf{a}_b + \boldsymbol{\omega}_{ib}^b\times(\boldsymbol{\omega}_{ib}^b\times\boldsymbol{\rho}_i) + \boldsymbol{\dot{\omega}}_{ib}^b\times\boldsymbol{\rho}_i - \mathbf{g}_{b}
\end{equation}

 Equation (\ref{eq:gf}) is the fundamental GF-equation, and serves as the basis for the equations of motion. Note, that while we measure $\mathbf{\hat{f}}_b$ directly and $\mathbf{g}_b$ can be calculated from known constants and the body's orientation, the calculation of $\boldsymbol{\omega}$,$\boldsymbol{\dot{\omega}}$ requires an estimation algorithm, typically a variation of a least-squares process, as can be seen in the following section. 

\subsection{Gyro-Free Variable Notation}
The literature provides an inconsistent notation of this variable, some common options include:
\begin{enumerate}
    \item Denoting the term $\mathbf{f}_{ib}^b$, which represents the value $\mathbf{a}_b-\mathbf{g}_b$.
    \item Explicitly writing $\mathbf{a}_b-\mathbf{g}_b$. Defining either one or both of them as variables later in the paper.
\end{enumerate}
In this paper we use the $\mathbf{a}_b-\mathbf{g}_b$ notation, while acknowledging the widespread use of the other options in the literature.
\subsection{Gyro-Free Least Squares Formulation \label{subsec:non-linear-lsq}}
We follow \cite{pachter2013gyro} for the presentation of the optimization equations expressed in an inertial frame. Looking at Eq. (\ref{eq:gf}), by assuming $\boldsymbol{\omega}_{ib}^b$ is received through integration, our variable set $\mathbf{x}$ is defined as:
\begin{equation}
    \mathbf{x}=\begin{bmatrix}\boldsymbol{\dot{\omega}_{ib}^b}&\mathbf{a}_b-\mathbf{g}_b\\
\end{bmatrix}^T
\end{equation}
Following this definition, a linear system is obtained by stacking Eq. (\ref{eq:gf}) N times, where N is the number of IMUs:
\begin{equation}
    \begin{bmatrix}
\mathbf{\hat{f}}_{b,\,1} \\
\vdots \\
\mathbf{\hat{f}}_{b,\,N}\end{bmatrix}= \begin{bmatrix}
\times\boldsymbol{\rho}_1 & 1 \\
\vdots  & \vdots \\
\times\boldsymbol{\rho}_N & 1 \\
\end{bmatrix}\mathbf{x} + \begin{bmatrix}
\boldsymbol{\omega}_{ib}^b\times(\boldsymbol{\omega}_{ib}^b\times\boldsymbol{\rho}_1) \\
\vdots \\
\boldsymbol{\omega}_{ib}^b\times(\boldsymbol{\omega}_{ib}^b\times\boldsymbol{\rho}_N) \\
\end{bmatrix}
\end{equation}
Let $\mathbf{H}$ be defined as:
\begin{equation}
\mathbf{H}= \begin{bmatrix}
\times\boldsymbol{\rho}_1 & 1 \\
\vdots  & \vdots \\
\times\boldsymbol{\rho}_N & 1 \\
\end{bmatrix}
\end{equation}
Then by defining the pseudo-inverse matrices $\mathbf{H}_{\boldsymbol{\dot{\omega}}},\,\mathbf{H}_{\mathbf{a-g}}$:
\begin{equation}
     \begin{bmatrix}\mathbf{H}_{\boldsymbol{\dot{\omega}}} \\ \mathbf{H}_{\mathbf{a-g}}\end{bmatrix} = (\mathbf{H}^T\mathbf{H})^{-1}\mathbf{H}^T
\end{equation}
A solution for the overdetermined system can be found:
\begin{equation}
    \label{eq:gf-solution-simple}
    \mathbf{x}=\begin{bmatrix}\mathbf{H}_{\boldsymbol{\dot{\omega}}} \\ \mathbf{H}_{\mathbf{a-g}}\end{bmatrix}\cdot(\mathbf{Y}-\mathbf{M})
\end{equation}
Where:
\begin{equation}
    \mathbf{Y}=\begin{bmatrix}
\mathbf{\hat{f}}_{b,\,1} \\
\vdots \\
\mathbf{\hat{f}}_{b,\,N}\end{bmatrix}
\end{equation}
\begin{equation}
    \mathbf{M}=\begin{bmatrix}
\boldsymbol{\omega}_{ib}^b\times(\boldsymbol{\omega}_{ib}^b\times\boldsymbol{\rho}_1) \\
\vdots \\
\boldsymbol{\omega}_{ib}^b\times(\boldsymbol{\omega}_{ib}^b\times\boldsymbol{\rho}_N) \\
\end{bmatrix}
\end{equation}
Solving Eq. (\ref{eq:gf-solution-simple}) provides the GF output for each MIMU measurement, if the following conditions are met:
\begin{enumerate}
    \item Initial conditions (orientation, angular velocity) are known.
    \item The matrix $\mathbf{H}^T\mathbf{H}$ is non-singular.
    \item $N\geq6$.
\end{enumerate}
\subsection{Gyro-Free Orientation Propagation}
We begin with a set of $\textrm{N}$ IMUs measurements in their respective sensor frame. Applying the constant sensor to body rotation $\mathbf{R}_{s_i}^b$ we are left with a set of \textbf{N} IMUs measurements in the body frame ($\mathbf{f}_{b_1},\dots,\mathbf{f}_{b_i},\dots,\mathbf{f}_{b_n}$).

Assuming the initial angular velocity and acceleration are known, we can derive $\mathbf{a}_b-\mathbf{g}_b,\,\boldsymbol{\dot{\omega}}_{ib}^b$ based on an estimation process based on (\ref{eq:gf}) as seen in section \ref{subsec:non-linear-lsq}. 

The rest of the process follows standard nonlinear INS kinematic equations, as covered extensively in \cite{titterton2004strapdown}, \cite{groves2008}. The body rate with respect to the navigation frame is:

\begin{equation}
    \label{eq:wnb}
    \boldsymbol{\omega}_{nb}^b=\boldsymbol{\omega}_{ib}^b-\mathbf{R}_b^{n\;-1}\cdot\boldsymbol{\omega}_{in}
\end{equation}

 Where $\boldsymbol{\omega}_{in}$ is the sum of the Earth's rate with respect to the inertial frame, $\boldsymbol{\omega}_{ie}$, and the turn rate of the navigation frame with respect to the Earth $\boldsymbol{\omega}_{en}$. Note that for many AHRS applications the body's position and velocity information, which are required to compute $\boldsymbol{\omega}_{ie}$, is not available and average or expected values are used instead.

The transformation matrix rate of change is given by:

\begin{equation}
    \label{eq:DCMdot}
    \mathbf{\dot{R}}_b^n = \mathbf{R}_b^n\cdot\boldsymbol{\Omega}_{nb}^b
\end{equation}

 Where $\boldsymbol{\Omega}_{nb}^b$ is the skew symmetric form of $\boldsymbol{\omega}_{nb}^b$. Finally, $\mathbf{R}_b^n$ is found by integrating (\ref{eq:DCMdot}).

\begin{equation}
    \label{eq:DCMint}
    \mathbf{R}_b^{n\,(k+1)} = \mathbf{R}_b^{n\,(k)}\cdot \exp(\mathbf{\dot{R}}_b^n \cdot \Delta t)
\end{equation}
\begin{figure*}[!ht]
    \centering
    \includegraphics[width=\textwidth]{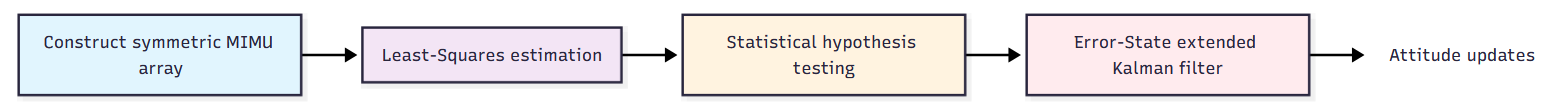}
    \caption{Flowchart of our proposed approach. After the symmetrical MIMU array is constructed, a least squares optimizer is used to evaluate the GF-output, which is then filtered using an hypothesis test before being fed into an ES-EKF to produce an attitude update.}
    \label{fig:flowchart_simple}
\end{figure*}
\subsection{Error-state AHRS Kalman Filter \label{subsec:ES-AHRS}}
Attitude and Heading Reference Systems (AHRS) are a subset of algorithms which handle the fusion of angular velocity, traditionally measured with a gyroscope, and specific force, measured from the accelerometer and optionally a heading measurement from a compass. 

In order to fuse multiple sensors, an estimation algorithm is employed, traditionally a Kalman Filter, under the assumption of independent process and measurement noises. Since the nature of the system is nonlinear, the nonlinear version of the Kalman Filter, named the Extended Kalman Filter (EKF) is used. In navigation, it is common to use the error state (ES) implementation, which evaluates the error in the solution, to correct the INS propagated solution \cite{farrell2008aided}. 

For the basic AHRS ES-EKF implementation, a three error state vector consisting of the attitude errors is implemented:

\begin{equation}
    \label{eq:dx}
    \boldsymbol{\delta}\mathbf{x}=\begin{bmatrix}\boldsymbol{\delta\Psi}\end{bmatrix}^T    
\end{equation}

 The linearized error state model is:
\begin{equation}
    \label{eq:linearized}
    \delta\dot{\mathbf{x}} = \mathbf{F}\delta\mathbf{x} + \mathbf{G}\mathbf{w}
\end{equation}

Where $\mathbf{F}$ is the system matrix, $\mathbf{G}$ is the shaping matrix, and $\mathbf{w}$ is the noise vector. When biases are not accounted for, the system matrix is defined as:

\begin{equation}
    \label{eq:F}
    \mathbf{F} = \mathbf{0}_{3}
  \end{equation}

 Similarly, the shaping matrix is given by:

\begin{equation}
    \label{eq:G}
    \mathbf{G}=\begin{bmatrix} -\mathbf{R}_b^n \end{bmatrix}
\end{equation}

 And the noise vector is:

\begin{equation}
    \label{eq:w}
    \mathbf{w} = \begin{bmatrix}\mathbf{w}_g\end{bmatrix}
\end{equation}

 Where $\mathbf{w}_g$ are the gyro zero-mean Gaussian white noise with a standard deviation of $\sigma_g$. Following from that, the process noise covariance matrix, $\mathbf{Q_w}$, is given by:

\begin{equation}
    \mathbf{Q_{w}} = \sigma_g^2 \cdot \mathbf{I}_3
\end{equation}

 In addition to the state model, the AHRS measurement model works under an assumption of momentarily zero-acceleration. In such event, we can deduce that:

\begin{equation}
    \label{eq:zero-acceleration}
    \mathbf{\hat{f}}_b \approx -\mathbf{\hat{g}}_b
\end{equation}
\begin{equation}
    \label{eq:dg}
    \boldsymbol{\delta}\mathbf{g} = \mathbf{g}_n - \mathbf{R}_b^{n\;-1}\cdot(-\mathbf{\hat{f}}_b)
\end{equation}

Which allows us to use the following measurement matrix:

\begin{equation}
    \label{eq:H}
    \mathbf{H}=\begin{bmatrix}-\mathbf{g}_n\times\end{bmatrix}
\end{equation}

With the measurement shaping matrix:

\begin{equation}
    \label{eq:n}
    \mathbf{n}=\begin{bmatrix} \mathbf{R}_b^n \end{bmatrix}
\end{equation}

Which results in the following measurement model:

\begin{equation}
    \label{eq:measurement}
    \delta{\mathbf{g}} = \mathbf{H}\delta\mathbf{x} + \mathbf{n}\boldsymbol{\nu}
\end{equation}

 Where $\boldsymbol{\nu}$ is the noise vector containing the accelerometer zero-mean Gaussian white noise with a standard deviation of $\sigma_f$ and a covariance matrix given by:

\begin{equation}
    \mathbf{Q_{\nu}} = \sigma_f^2 \cdot \mathbf{I}_3
\end{equation}

The rest of the ES-EKF algorithm is described by as follows:

\begin{enumerate}
  \item Use (\ref{eq:wnb})-(\ref{eq:DCMint}) to compute the attitude updates.
  \item Evaluate the system matrix $\mathbf{F}_k$ and the shaping matrix $\mathbf{G}_k$.
  \item Apply a first-order approximation to evaluate the transition matrix $\boldsymbol{\Phi}$:
  \begin{equation}
      \boldsymbol{\Phi}_k=\mathbf{I}_{3}+\mathbf{F}_k\cdot \Delta t.
  \end{equation}
  \item Propagate the error state covariance matrix, {$\mathbf{P}$}:
  \begin{equation}
      \mathbf{P}^-_{k} = \boldsymbol{\Phi}_k \mathbf{P}^+_{k - 1}\boldsymbol{\Phi}_k^T + \mathbf{G}\mathbf{Q_w}\mathbf{G}^T\cdot\Delta t.
  \end{equation} 
\end{enumerate}

Once a zero-acceleration measurement is detected, the error-state EKF correction equations are applied as follows:

\begin{enumerate}
    \item Use (\ref{eq:dg}) to compute the measurement residual
    \item Evaluate the measurement matrix, $\mathbf{H}_k$.
    \item Compute the Kalman gain, $\mathbf{K}_k$:
    \begin{equation}
        \mathbf{K}_k = \mathbf{P}_k\mathbf{H}_k^T\begin{pmatrix}\mathbf{H}_k\mathbf{P}_k\mathbf{H}_k^T + \boldsymbol{\nu}\mathbf{Q_{\nu}}\boldsymbol{\nu}^T\end{pmatrix}^{-1}
    \end{equation}
    \item Compute the a posteriori error state, $\boldsymbol{\delta}\mathbf{x}_k^+$:
    \begin{equation}
        \boldsymbol{\delta}\mathbf{x}_k^+ = \mathbf{K}_k\boldsymbol{\delta}\mathbf{g}_k.
    \end{equation}
    \item Correct the navigation solution $\mathbf{x}_k$:
    \begin{equation}
        \mathbf{x}^+_k = \mathbf{x}^-_k-\boldsymbol{\delta}\mathbf{x}_k^+.
    \end{equation}
    \item Update the error-state covariance matrix:
    \begin{equation}
        \mathbf{P}_k^+ = \begin{pmatrix}\mathbf{I}_{3} - \mathbf{K}_k\mathbf{H}_k\end{pmatrix}\mathbf{P}_k^-.
    \end{equation}
    \item Set $\boldsymbol{\delta}\mathbf{x}_k = 0$ for the closed loop mechanism.
\end{enumerate}
\section{Proposed Approach}
The goal of our approach is to reduce the error of the GF-process output. It contains four main steps. First, a MIMU array is constructed in symmetric pairs to separate linear and rotational acceleration components. Next, a least squares estimation process independently calculates these components, addressing correlation issues. An error-state EKF is then applied to control the orientation propagation and corrections. Finally, statistical hypothesis testing is used to detect and eliminate noise-based propagation, ensuring accurate attitude updates. We begin by presenting the limitations of current gyro free formulation followed by a detailed description of each step of our proposed algorithm.
\subsection{Limitation in the Gyro-Free formulation \label{subsec:limitation}}
Analyzing (\ref{eq:gf}) provides multiple insights into the inherent limitations of the gyro free mechanization and optimization process.
\begin{enumerate}
    \item The linear component from (\ref{eq:gf}) is more dominant in the least-squares process (\ref{eq:gf-solution-simple}), which results in statistical suppression of the angular component. This phenomena may lead to an incorrect estimation of the angular components, coupled with an overly-confidence variance estimation.
    \item In the formulation presented in (\ref{eq:gf-solution-simple}) the least-squares process solves for the linear component alongside the angular components, which leads to cross-correlation between them (i.e. the estimated least-squares covariance matrix contains off-diagonal elements). In the Kalman filter, it is assumed the process and measurement noises are uncorrelated. If they are, the cross-covariance matrix should be taking into account in the Kalman formulation. Yet. in real-world scenarios such correlation are challenging to model and compensate. Such situation occurs in GF situations if the linear component is used to update the prediction phase which is based on the angular components. \item Deriving $\mathbf{R}_b^n$ from $\boldsymbol{\dot{\omega}}$ requires double-integration, which increases the rate of divergence in the mechanization process.
\end{enumerate}
Our proposed approach resolves limitations (1), (2) by separating the linear and angular components estimation, while mitigating the effects of (3) by applying statistical hypothesis-tests.
\subsection{3D Symmetrical MIMU}
At the basis of the proposed approach lies a subclass of MIMU arrays which we will define as symmetrical MIMU (SMIMU) arrays. It is assumed that each IMU contains three perpendicular accelerometers. A MIMU array is symmetric if for each IMU at point i exist an IMU at point j which satisfies:

\begin{equation}
    \label{eq:symmmetric}
    \boldsymbol{\rho}_j=-\boldsymbol{\rho}_i
\end{equation}

 That is, all IMUs in the architecture could be grouped into pairs $\{i,j\}$ in which the two IMUs are located on the opposite ends of a 3D diagonal. For an arbitrary symmetrical pair satisfying (\ref{eq:symmmetric}), the specific vector measurement are:
\begin{equation}
    \label{eq:f_b_i}
    \mathbf{\hat{f}}_{b_i} = \mathbf{a}_b + \boldsymbol{\omega}_{ib}^b\times(\boldsymbol{\omega}_{ib}^b\times\boldsymbol{\rho}_i) + \boldsymbol{\dot{\omega}}_{ib}^b\times\boldsymbol{\rho}_i - \mathbf{g}_{b}
\end{equation}
\begin{equation}
    \label{eq:f_b_j}
    \mathbf{\hat{f}}_{b_j} = \mathbf{a}_b - \boldsymbol{\omega}_{ib}^b\times(\boldsymbol{\omega}_{ib}^b\times\boldsymbol{\rho}_i) - \boldsymbol{\dot{\omega}}_{ib}^b\times\boldsymbol{\rho}_i - \mathbf{g}_{b}
\end{equation}
where $\mathbf{\hat{f}}_{b_i}$, $\mathbf{\hat{f}}_{b_j}$ are the specific forced measured by the i'th and j'th accelerometer, respectively.
In this case, we define two measurements combinations $\mathbf{\bar{f}}, \mathbf{\breve{f}}$ to be:
\begin{equation}
    \label{eq:f_bar_base}
    \mathbf{\bar{f}}_{ij} = \frac{1}{2}\cdot(\mathbf{\hat{f}}_{b_i}+\mathbf{\hat{f}}_{b_j})
\end{equation}
\begin{equation}
    \label{eq:f_breve_base}
    \mathbf{\breve{f}}_{ij} = \frac{1}{2}\cdot(\mathbf{\hat{f}}_{b_i}-\mathbf{\hat{f}}_{b_j})
\end{equation}
Using (\ref{eq:f_b_i})-(\ref{eq:f_b_j}), we rewrite (\ref{eq:f_bar_base})-(\ref{eq:f_breve_base}) as:
\begin{equation}
    \label{eq:f_bar}
    \mathbf{\bar{f}}_{ij} = \mathbf{a}_b - \mathbf{g}_b
\end{equation}
\begin{equation}
    \label{eq:f_breve}
    \mathbf{\breve{f}}_{ij} = \boldsymbol{\omega}_{ib}^b\times(\boldsymbol{\omega}_{ib}^b\times\boldsymbol{\rho}_i) + \boldsymbol{\dot{\omega}}_{ib}^b\times\boldsymbol{\rho}_i
\end{equation}
Thus, by employing the properties of the SMIMU array, we have separated the measurement into its linear and rotational acceleration components, which can be estimated independently. We have therefore removed limitation (1) as defined in Section \ref{subsec:limitation}.

\subsection{2D Symmetrical MIMU \label{subseq:2d_sym}}
The requirement for full 3D symmetry may be quite strict for some applications. Therefore, we also derive the 2D SMIMU, assuming all IMUs lie on a 2D plane. Without the loss of generality,  we focus our derivation on the X-Y plane. The same procedure can be applied to any 2D plane. To this end, $\boldsymbol{\rho}_j$ is defined as:

\begin{equation}
    \label{eq:relaxed_rho}
    \boldsymbol{\rho}_j = [-\rho_{i_x}, -\rho_{i_y}, \rho_{i_z}]^T
\end{equation}

Substituting (\ref{eq:relaxed_rho}) into (\ref{eq:f_bar})-(\ref{eq:f_breve}) results in:

\begin{equation}
    \label{eq:system_smimu_relaxed_direct}
    \resizebox{.7\linewidth}{!}{$
    \begin{aligned}
            \bar{f}_x &= a_x + \omega_x\omega_z\rho_z + \dot{\omega}_y\rho_z - g_x \\
            \bar{f}_y &= a_y + \omega_y\omega_z\rho_z - \dot{\omega}_x\rho_z - g_y \\
            \bar{f}_z &= a_z - \omega_x^2\rho_z - \omega_y^2\rho_z - g_z \\
            \breve{f}_x &= -\omega_y^2\rho_x - \omega_z^2\rho_x + \omega_x\omega_y\rho_y - \dot{\omega}_z\rho_y \\
            \breve{f}_y &= -\omega_x^2\rho_y - \omega_z^2\rho_y + \omega_x\omega_y\rho_x + \dot{\omega}_z\rho_x \\
            \breve{f}_z &= \omega_x\omega_z\rho_x + \omega_y\omega_z\rho_y + \dot{\omega}_x\rho_y - \dot{\omega}_y\rho_x
    \end{aligned}
    $}
\end{equation}

As a result of the reduction to a 2D plane, the linear components of $\mathbf{\bar{f}}$ contains the angular effects from the reduced axis (here, the z-axis). Under the assumption that the single-axis rotational contribution is negligible when compared to the linear acceleration, we can further simplify (\ref{eq:system_smimu_relaxed_direct}) to:

\begin{equation}
    \label{eq:system_smimu_relaxed}
    \resizebox{.7\linewidth}{!}{$
    \begin{aligned}
            \bar{f}_x &= a_x - g_x \\
            \bar{f}_y &= a_y - g_y \\
            \bar{f}_z &= a_z - g_z \\
            \breve{f}_x &= -\omega_y^2\rho_x - \omega_z^2\rho_x + \omega_x\omega_y\rho_y - \dot{\omega}_z\rho_y \\
            \breve{f}_y &= -\omega_x^2\rho_y - \omega_z^2\rho_y + \omega_x\omega_y\rho_x + \dot{\omega}_z\rho_x \\
            \breve{f}_z &= \omega_x\omega_z\rho_x + \omega_y\omega_z\rho_y + \dot{\omega}_x\rho_y - \dot{\omega}_y\rho_x
    \end{aligned}
    $}
\end{equation}

 The relaxed formulation contains all the elements of $\boldsymbol{\omega},\boldsymbol{\dot{\omega}}$. However, we can observe that $\dot{\omega}_x,\dot{\omega}_y$ are now linearly dependent since they are both observed only though the measurement $\breve{f}_z$.
\subsection{Symmetrical GF IMU \label{subsec:non-linear-lsq-new}}
The separation of the linear and rotational acceleration comes with one additional practical benefit. The condition for entering the AHRS measurement phase is that the body acceleration is approximately zero ($\mathbf{a}_n \approx 0$). Therefore, by evaluating the linear component only in the measurement phase the least squares solution of (\ref{eq:f_bar}) is given by:

\begin{equation}
    \begin{aligned}
        \label{eq:g_b_lstsq}
        \mathbf{\hat{g}}_b &= \frac{\Sigma_{i=1}^N -\mathbf{\bar{f}}_i}{N} \\
        \hat{\sigma}_{\mathbf{g}_b}^2 &= \frac{1}{N}\cdot\sigma^2_{\mathbf{\bar{f}}}
    \end{aligned}
\end{equation}

 To solve for $\boldsymbol{\omega}$ and $\boldsymbol{\dot{\omega}}$ from (\ref{eq:f_breve}), an iterative least-squares solution. The steps described here assume 3D SMIMU configuration, however the same steps can be applied for the 2D case.
\newline
Expanding the rotational equation (\ref{eq:f_breve}) we get:

\begin{equation}
    \label{eq:s_gf_system}
    \resizebox{.85\linewidth}{!}{$
    \begin{aligned}
            \breve{f}_x &= -\omega_y^2\rho_x - \omega_z^2\rho_x + \omega_x\omega_y\rho_y + \omega_x\omega_z\rho_z + \dot{\omega}_y\rho_z - \dot{\omega}_z\rho_y \\
            \breve{f}_y &= -\omega_x^2\rho_y - \omega_z^2\rho_y + \omega_x\omega_y\rho_x + \omega_y\omega_z\rho_z + \dot{\omega}_z\rho_x - \dot{\omega}_x\rho_z \\
            \breve{f}_z &= -\omega_x^2\rho_z - \omega_y^2\rho_z + \omega_x\omega_z\rho_x + \omega_y\omega_z\rho_y + \dot{\omega}_x\rho_y - \dot{\omega}_y\rho_x
    \end{aligned}
    $}
\end{equation}

 Since $\boldsymbol{\rho}$ is known, the variable vector, $\mathbf{x}_{SGF}$, is defined as:

\begin{equation}
    \mathbf{x}_{SGF} = \begin{bmatrix}\boldsymbol{\omega}, \boldsymbol{\dot{\omega}}
    \end{bmatrix}^T
\end{equation}

The Jacobian $\mathbf{A}_{SGF}$ is defined as follows:

\begin{equation}
    \resizebox{.85\linewidth}{!}{$
    \mathbf{A}_{SGF} = \begin{bmatrix}
        \frac{\partial f_x}{\partial \omega_x} & \frac{\partial f_x}{\partial \omega_y} & \frac{\partial f_x}{\partial \omega_z} & \frac{\partial f_x}{\partial \dot{\omega}_x} & \frac{\partial f_x}{\partial \dot{\omega}_y} & \frac{\partial f_x}{\partial \dot{\omega}_z} \\
        \frac{\partial f_y}{\partial \omega_x} & \frac{\partial f_y}{\partial \omega_y} & \frac{\partial f_y}{\partial \omega_z} & \frac{\partial f_y}{\partial \dot{\omega}_x} & \frac{\partial f_y}{\partial \dot{\omega}_y} & \frac{\partial f_y}{\partial \dot{\omega}_z} \\
        \frac{\partial f_z}{\partial \omega_x} & \frac{\partial f_z}{\partial \omega_y} & \frac{\partial f_z}{\partial \omega_z} & \frac{\partial f_z}{\partial \dot{\omega}_x} & \frac{\partial f_z}{\partial \dot{\omega}_y} & \frac{\partial f_z}{\partial \dot{\omega}_z}
        \end{bmatrix}
        $}
\end{equation}

Evaluating the partial derivatives, we arrive at:

\begin{equation}
     \mathbf{A}_{SGF}=\begin{bmatrix}\mathbf{B} & [\boldsymbol{\rho}\times] \end{bmatrix}
\end{equation}

where $\mathbf{B}$ is defined by:
 \begin{equation}
    \label{eq:B_matrix}
    \resizebox{.85\linewidth}{!}{$
    \mathbf{B} = \begin{bmatrix}
    \omega_y\rho_y + \omega_z\rho_z & -2\omega_y\rho_x + \omega_x\rho_y & -2\omega_z\rho_x + \omega_x\rho_z  \\
    -2\omega_x\rho_y + \omega_y\rho_x & \omega_x\rho_x + \omega_z\rho_z & -2\omega_z\rho_y + \omega_y\rho_z & \\
    -2\omega_x\rho_z + \omega_z\rho_x & -2\omega_y\rho_z + \omega_z\rho_y & \omega_x\rho_x + \omega_y\rho_y
    \end{bmatrix}
    $}
\end{equation}

The measurement residual is defined as:

\begin{equation}
    \mathbf{v}^{(k)}_{SGF} = \mathbf{\hat{l}}^{(k)}_{SGF} - \mathbf{l}_{SGF}
\end{equation}

where $\mathbf{\hat{l}}_{SGF}$ is the measurement vector, as defined in 
\begin{equation}
    \mathbf{l}_{SGF} = \begin{bmatrix} \dots ,\breve{f}_{x_i},\breve{f}_{y_i},\breve{f}_{z_i},\dots     
    \end{bmatrix} ^ T
\end{equation}

 Finally, the measurement covariance matrix, $\mathbf{P}_{\mathbf{\breve{f}}}$, is defined as:

\begin{equation}
    \mathbf{P}_{\mathbf{\breve{f}}} = \sigma_{\breve{f}}\cdot\mathbf{I}_3 
\end{equation}

 Given an initial values for $\boldsymbol{\omega}^{(0)}$, we evaluate $\mathbf{A}_{SGF}^{(0)}$ (by evaluating $\mathbf{B}^{(0)}$) and iteratively improve our estimation by applying the step function:
 \begin{equation}
    \label{eq:step_function}
    \mathbf{x}^{(k+1)} = \mathbf{x}^{(k)} + (\mathbf{A}^{(k)\,T}\mathbf{P}^{-1}\mathbf{A}^{(k)})^{-1}(\mathbf{A}^{(k)\,T}\mathbf{P}^{-1}\mathbf{v}_{SFG}^{(k)})
\end{equation}

Finally, the variable covariance matrix is calculated by:
\begin{equation}
    \label{eq:step_covariance}
        \mathbf{P}_{\mathbf{x}} =  \begin{bmatrix}
        \mathbf{P}_{\boldsymbol{\omega}} & \mathbf{P}_{\boldsymbol{\omega}\boldsymbol{\dot{\omega}}} \\
        \mathbf{P}_{\boldsymbol{\dot{\omega}}\boldsymbol{\omega}} & \mathbf{P}_{\boldsymbol{\dot{\omega}}} \\
        \end{bmatrix} = \boldsymbol{\sigma_\mathbf{v}}^2 (\mathbf{A}_{SGF}^{T}\,\mathbf{P}_{\mathbf{\breve{f}}}^{-1}\,\mathbf{A}_{SGF})^{-1}
\end{equation}

where $\boldsymbol{\sigma}^2_\mathbf{v}$ is:
\begin{equation}
    \boldsymbol{\sigma_\mathbf{v}}^{(k)\,2} = \frac{\mathbf{v}_{SFG}^{(k)\,T}\,\mathbf{P}_{\mathbf{\breve{f}}}^{-1}\,\mathbf{v}_{SFG}^{(k)}}{r}
\end{equation}

and $r$ is the degrees of freedom given by:
\begin{equation}
    r = \frac{3}{2}N-6
\end{equation}

Thus, the solution of the angular velocity vector and angular acceleration vector in nearly zero linear acceleration is given in (\ref{eq:step_function}) with the associated covariance (\ref{eq:step_covariance}).

\subsection{Measurement phase in Symmetrical MIMU Arrays}
When considering the application of the 3D SMIMU as part of the EKF measurement model (\ref{eq:measurement}), we ought to examine the transformation from $\{\mathbf{f}_i$, $\mathbf{f}_j\}$ to $\{\mathbf{\hat{f}}_{ij}$, $\mathbf{\breve{f}}_{ij}\}$. 

To this end, we write (\ref{eq:f_bar}) and (\ref{eq:f_breve}) as a system of equation for an arbitrary axis:

\begin{equation}
    \label{eq:gf_transform}
     \begin{bmatrix}\mathbf{\hat{f}}\\ \mathbf{\breve{f}}\end{bmatrix}_{ij} = \frac{1}{2}\cdot\begin{bmatrix} 1 & 1 \\ 1 & -1 \\ \end{bmatrix}\begin{bmatrix}\mathbf{f}_i\\ \mathbf{f}_j\end{bmatrix}
\end{equation}

Equation (\ref{eq:gf_transform}) is a linear system characterized by the transformation matrix, $\mathbf{T}$:

\begin{equation}
    \label{eq:t_mat}
    \mathbf{T} = \begin{bmatrix} \frac{1}{2} & \frac{1}{2} \\ \frac{1}{2} & -\frac{1}{2} \end{bmatrix}
\end{equation}

For any given IMU, the variance of an accelerometer measurement is:
\begin{equation}
    \label{eq:imu_sig}
     \mathbf{P}=\sigma^2
\end{equation}

Using (\ref{eq:imu_sig}) for IMUs' $\{i,j\}$, the combined measurement covariance matrix is:

\begin{equation}
    \mathbf{P}_{ij} = \begin{bmatrix} \sigma_i^2 & 0 \\ 0 & \sigma_j^2 \end{bmatrix}
\end{equation}

Thus, the transformed covariance matrix is:
\begin{equation}
    \label{eq:tpt_full}
    \mathbf{P}_{SMIMU}=\mathbf{T}\mathbf{P} \mathbf{T}^T = \begin{bmatrix} \frac{\sigma_i^2 + \sigma_j^2}{4} & \frac{\sigma_i^2 - \sigma_j^2}{4} \\ \frac{\sigma_i^2 - \sigma_j^2}{4} & \frac{\sigma_i^2 + \sigma_j^2}{4} \end{bmatrix}
\end{equation}

 As the IMUs are of a similar grade, $\boldsymbol{\sigma}_i=\boldsymbol{\sigma}_j$, (\ref{eq:tpt_full}) is simplified to:

\begin{equation}
    \label{eq:s_cov}
    \mathbf{P}_{SMIMU}=\begin{bmatrix} \sigma^2_{\hat{f}}& 0 \\ 0 & \sigma^2_{\breve{f}} \\
\end{bmatrix} = \frac{1}{2}\sigma_i^2\cdot\mathbf{I}_2
\end{equation}

 Equation (\ref{eq:s_cov}) provides an important insight - the MIMU symmetrical transformation is orthogonal (i.e. its results remain uncorrelated). Since we can separately the estimation of the linear and rotation parts, the symmetrical formulation had fulfilled the requirements of the Kalman filter (uncorrelated measurement and process noise). Therefore, we resolved limitation (2) as defined in Section \ref{subsec:limitation} and we can apply the error-state EKF as described in Section \ref{subsec:ES-AHRS}, or any other implementation of the AHRS-EKF.

\subsection{Hypothesis Testing}
The GF-IMU diverges quicker when compared to its standard-IMU counterpart due to the double integration of the angular acceleration. In some situations, this solution could fully diverge under certain conditions. We therefore look to mitigate the instability by reducing the number of unnecessary propagations (i.e., propagation which is a result of noise rather than actual measurements).

Returning to (\ref{eq:step_covariance} we observe that the least-squares estimated covariance contains the angular velocity covariance matrix $\mathbf{P}_{\boldsymbol{\omega}}$. Therefore, we can construct an hypothesis test on the angular velocity that checks if the measured amount is significant enough to be considered a measurement, given the known uncertainties. The hypothesis test is shown in Algorithm \ref{alg:test}.

\begin{algorithm}
    \caption{Hypothesis Testing on $\boldsymbol{\omega}$}
    \label{alg:test}
    \begin{algorithmic}[]
        \State Let $\alpha_c$ be the confidence coefficient 
        \State Given $\omega_x, \omega_y, \omega_z$ from (\ref{eq:step_function}) and $\mathbf{P}_{\boldsymbol{\omega}}$ from (\ref{eq:step_covariance})
        \State Calculate $\sigma_{\omega_x}, \sigma_{\omega_y}, \sigma_{\omega_z}$ from $\mathbf{P}_{\boldsymbol{\omega}}$
        \If{$\omega_i \leq \alpha_c\cdot\sigma_{\omega_i}, \; Given \; i\in[x:0,y:1,z:2]$}
            \State $\omega_i = 0$
            \For{$j\in[0,2]$}
                \State $\mathbf{P}_{\boldsymbol{\omega}}[i, j]=0$
            \EndFor
        \EndIf
\end{algorithmic}
\end{algorithm}

After performing the significance test, we set the variance of the tested parameter and its covariances with other parameters to zero. The reasoning behind this is as follows:

\begin{itemize}
    \item The goal of an hypothesis test is to resolve ambiguities (up to a certain confidence level, in this example 90\%). If as a result of a test we have set $\omega_i=0$, we have resolved its values and therefore it is no longer a random variable, as such its variance is zero.
    \item Since we set the value of $\omega_i$, any change in $\omega_j \; (j\neq i)$ will not effect it, therefore by definition the covariance between them is zero.
\end{itemize}

By applying the hypothesis test, unnecessary propagations are greatly reduced, which mitigate issue (3) as defined in Section \ref{subsec:limitation}.

\subsection{Summary}
\begin{figure*}[th!]
    \centering
    \includegraphics[width=\textwidth]{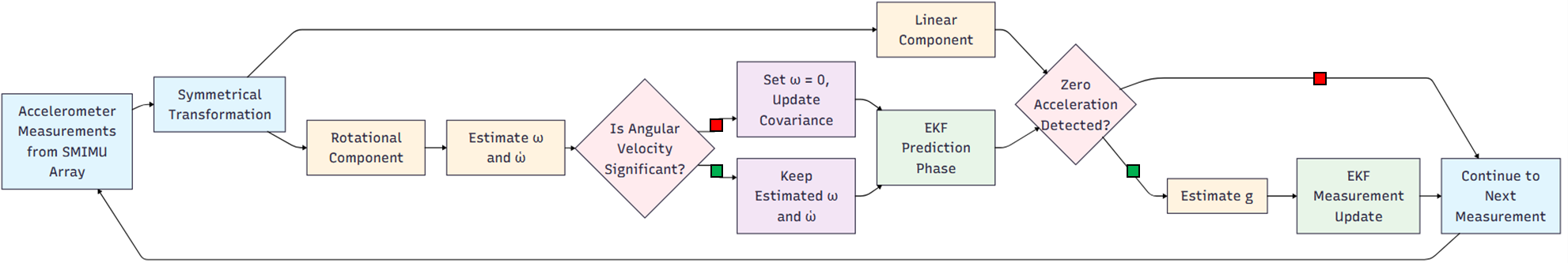}
    \caption{Proposed approach block diagram. The process begins with accelerometer measurements that undergo symmetrical transformation to differentiate between linear and rotational components. The rotational component is fed into the GF-least-squares estimation process, which alongside evaluating the angular velocity and acceleration calculates their standard deviations, which are used in the hypothesis test. If significant, the system proceeds through EKF prediction and if the body is not accelerating a measurement update phases.}
    \label{fig:flowchart}
\end{figure*}

Figure \ref{fig:flowchart} illustrates our proposed approach. It begins by applying the symmetrical transformation on each IMU pair to differentiate between linear and angular components. It then feeds the angular component into the GF angular least-squares optimization to calculate the angular velocity, acceleration and and related covariance matrices. When angular velocity is insignificant, it sets to zero by the hypothesis test, which then moves on to the AHRS Kalman filter prediction phase. The algorithm then assesses whether the platform is in steady-state conditions ($|\mathbf{\bar{f}}|\approx g_e$), and if detected, it applies the GF linear least-squares solution to evaluate the linear component, which is then fed to the Kalman filter measurement phase, producing updated state estimates and covariance matrices that serve as inputs for the next cycle. Thereby the proposed approach provides a continuous and accurate estimates of the platform's pitch and roll angles, with a continuous, yet diverging, estimation of the yaw angle. 

The proposed approach is presented here in the DCM formulation, but is applicable to other representations (e.g. Euler angles, Quaternions) as well.
\section{Analysis and Results}
\subsection{Datasets}
In this work, we utilized two open datasets to evaluate the performance of the proposed approach.
The first dataset, "Quadrotor Dead Reckoning with Multiple Inertial Sensors" by Hurwitz et al. \cite{hurwitz2023quadrotor} provides recordings from a DJI Phantom 4 RTK \cite{DJIRTK} drone. It was equipped with four Xsens DOT IMUs \cite{XDOT}, arranged around the RTK GNSS receiver, which serves as the dataset's ground truth (GT). For the construction of the 2D symmetrical arrays, IMUs $\{1,3\}$ and $\{2,4\}$ where joined as pairs.
The second dataset, "Multiple and Gyro-Free Inertial Dataset" (MAGF-ID) by Yampolsky et al. \cite{yampolsky2024multiple} provides recordings from both a car and a mobile robot (ROSbot), with the MRU-P \cite{InertialLabs} serving as the dataset's GT. For the construction of the 3D symmetrical arrays, IMUs $\{12,9\}$, $\{5,11\}$, $\{7,15\}$ and $\{8,10\}$ where joined as pairs. The quadcopter and car recordings were employed in this work. Table \ref{tab:sensor_specs} provides the sensor specification and Table \ref{tab:trajectories} list the available trajectories for each platform.

\begin{table}[htbp]
\centering
\caption{Sensor Specifications}
\label{tab:sensor_specs}
\begin{tabular}{lccc}
\toprule
\textbf{Parameter} & \textbf{Xsens DOT} & \textbf{MRU-P} & \textbf{Units} \\
\midrule
Sampling Rate & 120 & 100 & Hz \\
Accelerometer Bias Stability & 0.03 & 0.005 & mg \\
Gyroscope Bias Stability & 10 & 1 & °/hour \\
Accelerometer Noise Density & 120 & 0.025 & $\mu$g/$\sqrt{\text{Hz}}$ \\
Gyroscope Noise Density & 0.007 & 0.004 & °/s/$\sqrt{\text{Hz}}$ \\
\bottomrule
\end{tabular}
\end{table}

\begin{table}[htbp]
\centering
\caption{Dataset Trajectories}
\label{tab:trajectories}
\begin{tabular}{ccc}
\toprule
\textbf{Platform} & \textbf{Trajectory Type} & \textbf{Duration [min]} \\
\midrule
Quadrotor & Horizontal periodic & 16.1 \\
          & Vertical periodic & 12.6 \\
          & Straight line & 1.1 \\
\midrule
Car       & MP Square & 7.14 \\
          & MP to Bloom & 9.2 \\
          & Bloom Square & 9.4 \\
          & Bloom LM & 9.87 \\
          & Bloom Double LM & 10.63 \\
          & Bloom to MP & 9.1 \\
\midrule
\multicolumn{2}{c}{\textbf{Total Time}} & \textbf{85.14} \\
\bottomrule
\end{tabular}
\end{table}

\subsection{Performance Metrics}
To evaluate the performance of the proposed approach we employ the root mean square error (RMSE) performance metric. For attitude estimation, RMSE is calculated as:
\begin{equation}
RMSE = \sqrt{\frac{1}{N}\sum_{i=1}^{N}(\hat{\alpha}_i - \alpha_i)^2}
\end{equation}

where $\hat{\alpha}_i$ is the estimated angle (roll, pitch, or yaw) at time step $i$ and $\alpha_i$ is the corresponding GT. For comparative analysis, we compute both absolute RMSE and relative RMSE improvement. The absolute reduction is defined by:
\begin{equation}
\Delta RMSE_{abs} = RMSE_{baseline} - RMSE_{proposed}
\end{equation}

and the relative reduction is expressed as a percentage:
\begin{equation}
\Delta RMSE_{rel} = \frac{\Delta RMSE_{abs}}{RMSE_{baseline}} \times 100\%
\end{equation}

\subsection{Experiment Results}
\begin{table*}[!h]
\centering
\caption{RMSE results for the single IMU (baseline) and our SMIMU for the quadrotor and car datasets.}
\label{tab:attitude_results}
\scriptsize
\begin{tabular}{llcccc}
\toprule
\textbf{Platform} & \textbf{Trajectory}  & \textbf{Single IMU RMSE [deg]} & \textbf{GF RMSE [deg]} & \textbf{SMIMU RMSE [deg] (ours)} & \textbf{$\Delta\mathbf{RMSE}$ [deg/\%]}\\
\midrule
Quadrotor & Horizontal periodic & 4.55 & 3.61 & 2.90 & 1.6 (36\%) \\
          & Vertical periodic & 4.64 & 3.37 & 3.15 & 1.5 (32\%) \\
          & Straight line & 4.41 & 3.04 & 3.11 & 1.3 (29\%) \\
          & \textbf{Average} & \textbf{4.53} & \textbf{3.36} & \textbf{3.05} & \textbf{1.5 (32\%)} \\
\midrule
Car       & MP Square (Multi-Purpose parking lot) & 5.67 & 4.42 & 4.04 & 1.6 (29\%) \\
          & MP to Bloom (driving between parking lots) & 4.44 & 3.75 & 2.74 & 1.7 (38\%) \\
          & Bloom Square (Bloom parking lot) & 3.87 & 2.84 & 2.92 & 1.0 (25\%) \\
          & Bloom LM (Bloom parking lot lawn mower pattern) & 3.61 & 2.49 & 2.78 & 0.8 (23\%) \\
          & Bloom Double LM (double lawn mower pattern) & 5.67 & 5.79 & 4.04 & 1.6 (29\%) \\
          & Bloom to MP (driving between parking lots) & 4.87 & 3.91 & 3.03 & 1.8 (38\%) \\
          & \textbf{Average} & \textbf{4.69} & \textbf{3.53} & \textbf{3.26} & \textbf{1.42 (30\%)} \\
\bottomrule
\end{tabular}
\end{table*}
We compare between three approaches: single IMU , GF-IMU, and our proposed SMIMU. All of the approaches were processed through an ES-EKF AHRS. For the evaluations we employed all the trajectories as described in Table \ref{tab:trajectories}.
\subsubsection{\textbf{Attitude angles}}
\begin{figure}[!htbp]
     \centering
     \includegraphics[width=1\linewidth]{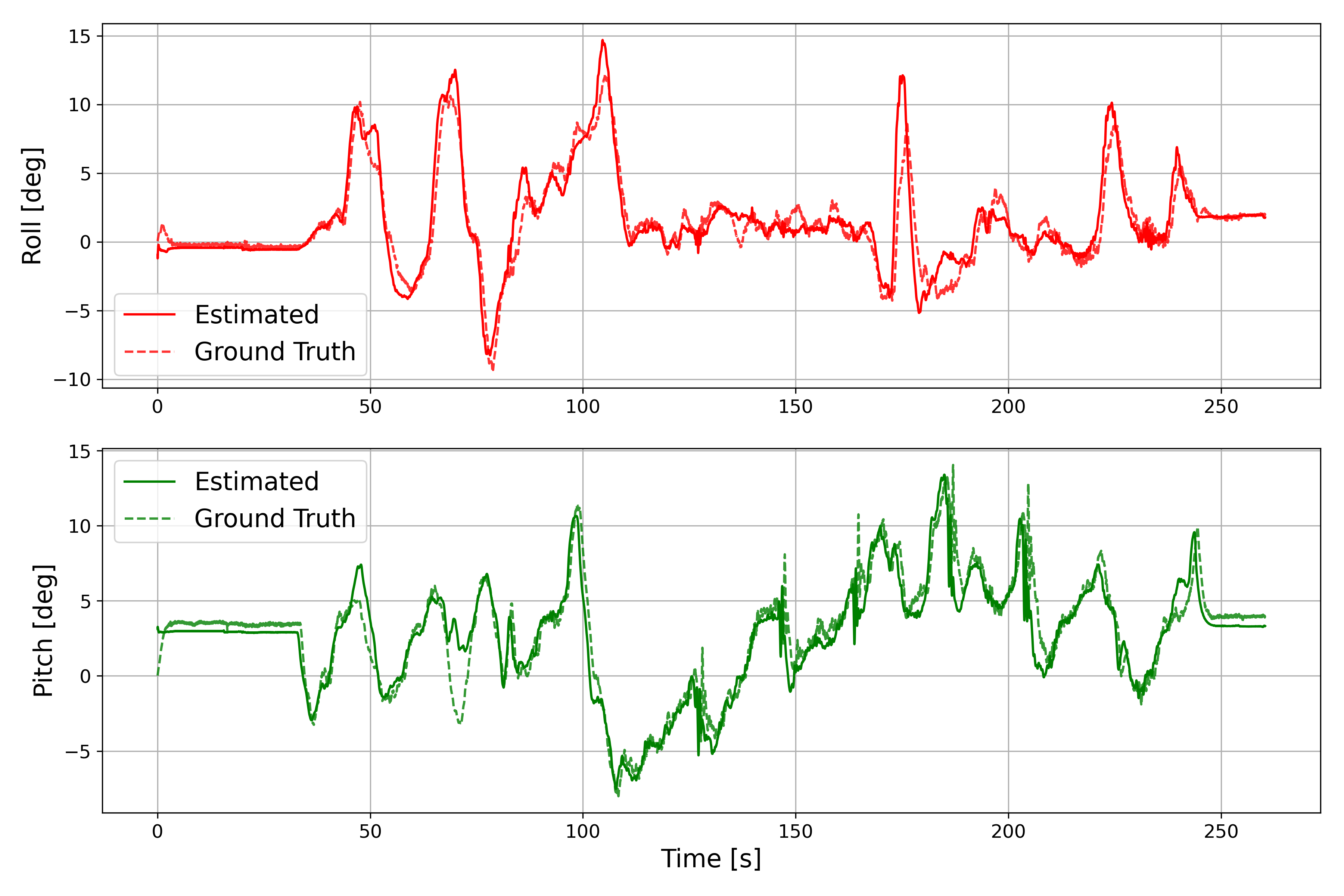}
     \caption{Comparison between ground truth and estimated roll/pitch angles for the Bloom to MP 1 trajectory.}
     \label{fig:pitch_roll_compare}
 \end{figure}

\begin{figure}[!htbp]
     \centering
     \includegraphics[width=1\linewidth]{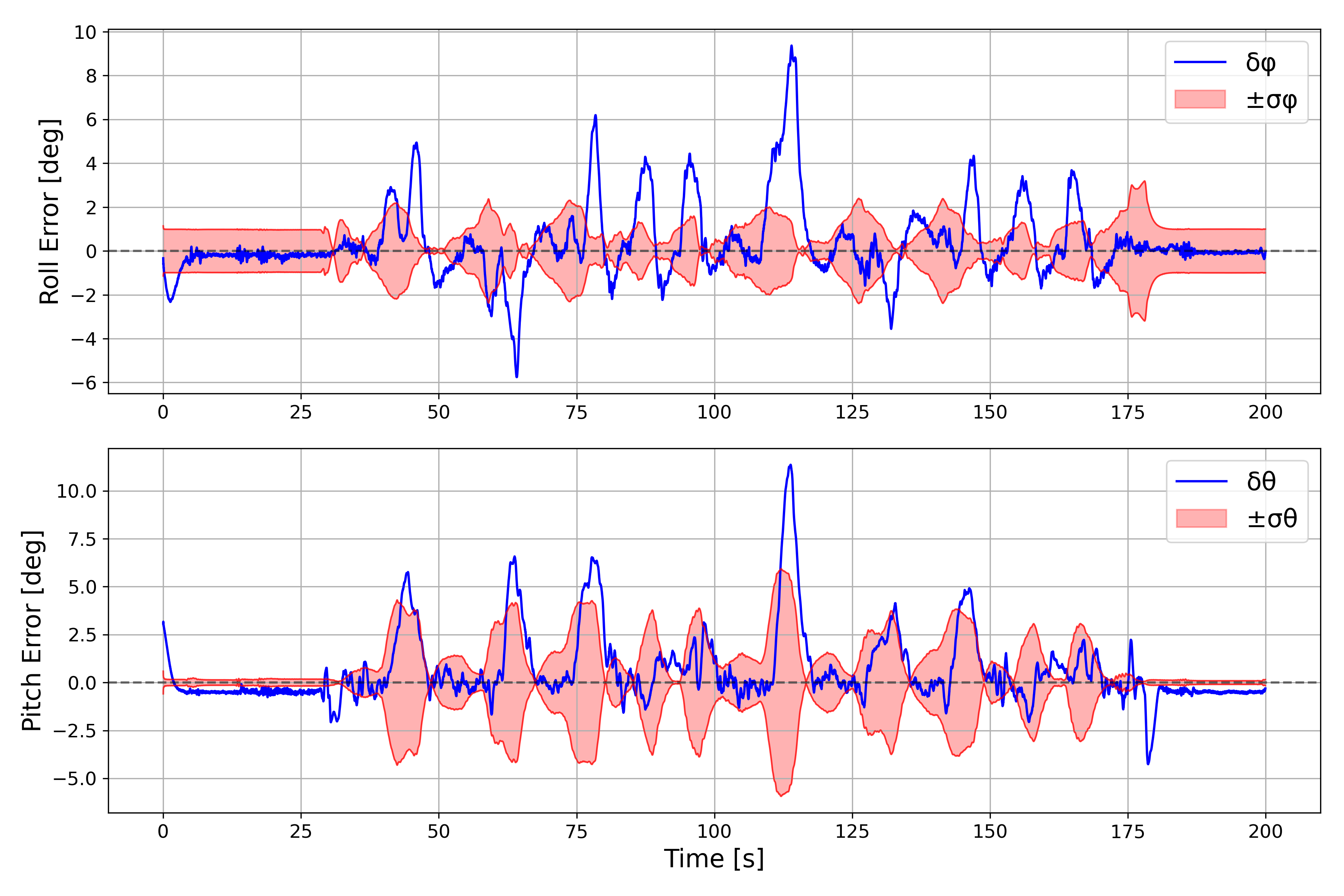}
     \caption{Covariance plot for the roll/pitch angles for the Bloom to LM 2 trajectory.}
     \label{fig:pitch_roll_covariance}
 \end{figure}

Table \ref{tab:attitude_results} summarizes the attitude estimation results. As can be seen for both platforms a significant and consistent improvement was demonstrated, ranging from 0.8 [deg] to 1.8 [deg] RMSE reduction (1.4 [deg] average reduction) and between 23-38\% relative RMSE improvement (30\% average improvement). Fig. \ref{fig:pitch_roll_compare} provides an example track for roll/pitch angle, as can be seen, while some under/over shooting is experienced especially when experiencing rapid attitude changes, the general estimated shape follows the GT angles closely. Likewise, Fig. \ref{fig:pitch_roll_covariance} provides an example for a covariance plot. It presents a stable and proper covariance behavior, diverging at the proper rate when no accelerometer update is available and converging when an updated is provided, with an estimated covariance ($1\;\sigma$) covering the expected $\sim68\%$ of the calculated error.

\subsubsection{\textbf{Heading angle}}
\begin{figure}[!htbp]
     \centering
     \includegraphics[width=1\linewidth]{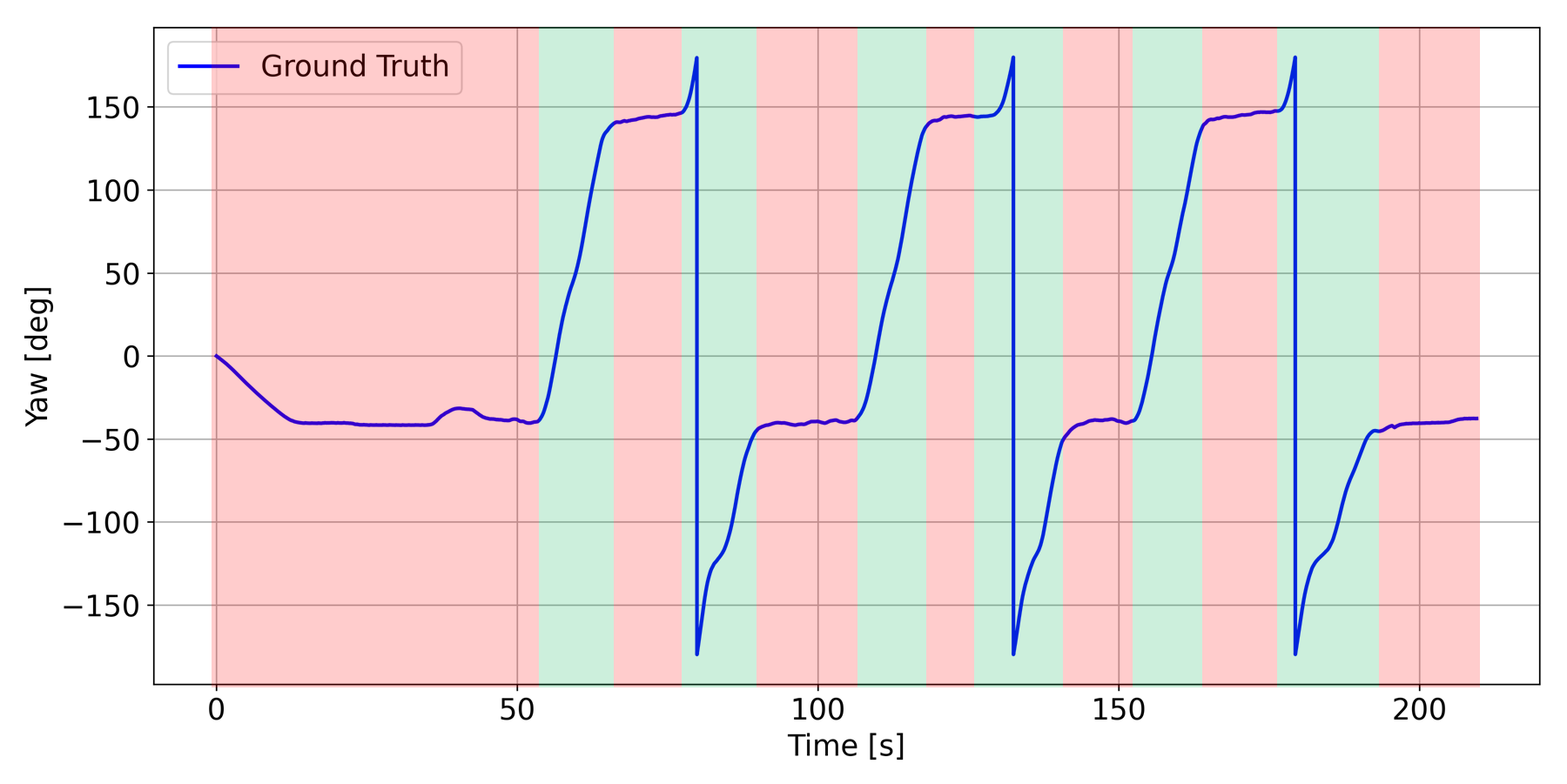}
     \caption{Yaw detection plot, MP Square 3. The red regions are where the hypothesis test failed (static) and the green regions are where the test passed (heading change)}
     \label{fig:yaw_detection}
 \end{figure}

Since the heading angle is not observeable, both the single IMU and the proposed approach diverge. However, our approach provides two additional benefits over the standard approach. To demonstrate them, Fig. \ref{fig:yaw_detection} provides a rotation detection graph of the heading angle. By employing the information provided by the hypothesis test, our approach successfully detects rotations (i.e. yaw changes) with more than 95\% accuracy. The algorithm's detections can be used in the logic of information-based filters (e.g. \cite{klein2020squeezing}) which relay upon accurate rotation detection for generating filter updates. In addition, since our estimation is both less noisy when compared to the single-IMU solution, and called upon less frequently since when a rotation is not detection the yaw stays constant, the divergence rate reduces significantly when compared to the state of the art. The reduction observed by our technique highly dependent on vehicle dynamics, with a greater improvement seen in trajectories with little to no rotations such as the quadrotor straight line trajectory where less than 0.5 [deg] drift was observed.

\section{Conclusions}
In this paper we introduce our novel symmetrical MIMU (SMIMU) array configuration. The proposed approach enables GF attitude estimation by the separation of linear and rotational components. Independently estimating these components in combination with the application of hypothesis testing and filtering within the AHRS GF EKF allows us to address the fundamental limitations of the standard GF implementation. Those include the suppression of the angular components in the estimation process, the correlation between the process and measurement noise, and mitigating the effects of double-integration on divergence rate by avoiding noise-based propagation. The proposed approach was validated using 85-minute dataset including recordings from both airborne and land platforms. Our SMIMU approach demonstrated a 30\% average RMSE improvement in attitude estimation compared to single-IMU baselines. In addition, we show that SMIMU achieves over 95\% rotation detection accuracy and reduced heading drift. These results enable reliable gyro-free navigation in applications where gyroscopes are unavailable, unreliable, or energy-constrained. Examples include miniature platforms, medical robotics, computational-constraint platforms, and long-endurance marine platforms (e.g. buoys).
%
\newpage
\bibliographystyle{IEEEtran}
\bibliography{export}
\end{document}